# Strain Tunable Band-gaps Of Two-dimensional Hexagonal BN And AlN: An FP-(L)APW+lo Study

Harihar Behera and Gautam Mukhopadhyay[*]

*Indian Institute of Technology Bombay, Powai, Mumbai-400076, India*
[*]*Email: gmukh@phy.iitb.ac.in* (Email of corresponding author)

**Abstract.** Using full potential density functional calculations within local density approximation (LDA), we found strain tunable band gaps of two-dimensional (2D) hexagonal BN (h-BN) and AlN (h-AlN) by application of in-plane homogeneous biaxial strain. The direct band gap of 2D h-BN turns indirect for compressive strains below 1.53% and remains direct under tensile strains up to 10%. However, the band gap of 2D h-AlN remains indirect for strains up to ±10%. While our result on 2D h-BN corroborates the reported strain effect on 2D h-BN (based on pseudo-potential method), our result on the strain tunable band gap of 2D h-AlN is something new. These results may find application in fabrication of future nano-electromechanical systems (NEMS) based on 2D h-BN and h-AlN.

**Keywords:** 2D h-AlN and h-BN, electronic structure, band-gap engineering, strain-tunable bang-gap, NEMS.
**PACS:** 63.22.Np; 73.61.Ey; 73.63.Bd; 73.22.-f

## INTRODUCTION

Nano-materials that can change energy gaps under mechanical strain are desirable for fabrication of nano-electromechanical systems (NEMS). With the recent synthesis [1-3] of 2D h-BN (so-called "white graphene"), comprising of alternating B and N atoms in a honeycomb lattice, the study of this material assumes much significance at time when this material is currently emerging as a promising substrate/gate dielectric for high-quality graphene electronics [4-5]. Because of its wide direct band gap [6] ($E_g$ = 5.971 eV), h-BN is seen as a promising material for ultraviolet laser devices [6-7]. On the other hand, there are reports of the synthesis of hexagonal AlN nanobelts [8], serrated nanoribbons [9]. Theoretical studies [10-11] predict stable graphene-like 2D hexagonal structures of AlN, which is considered recently [11] as an adequate template and/or gate insulator for silicene (the silicon analogue of graphene) [10-12]. Using first principles pseudo-potential method, J. Li et al. [13], reported strain induced (remarkable) modifications (such as the transition from direct to indirect band gap) of the band gap of 2D h-BN. However, the effect of strain on the band structures of 2D h-AlN is not reported yet.

Here, we report our simulated study on the effect of biaxial strain (which mimics the experimental situation when the material in question is supported on a stretchable substrate) on the band gaps of 2D h-BN and h-AlN, using the density functional theory (DFT) based full potential (linearized) augmented plane wave plus local orbital (FP-(L)APW+lo) method [14].

## CALCULATION METHODS

The calculations have been performed by using the Perdew-Zunger variant of LDA [15] and FP-(L)APW+lo method [14] as implemented in the elk-code [16]. The plane wave cutoff of $|\mathbf{G+k}|_{max}$ = 8.5/$R_{mt}$ (a.u.$^{-1}$) ($R_{mt}$ is the smallest muffin-tin radius in the unit cell) was used for the plane wave expansion of the wave function in the interstitial region. The k point mesh of (30×30×1) was used for all calculations. The total energy convergence was 2.0 μeV/atom between two successive steps. The 2D h-BN and h-AlN structures were simulated using three-dimensional hexagonal super-cells with a large value of the "$c$" ($c$ = 40 a.u.) parameter. The application of in-plane biaxial strain up to ±10% was simulated by varying the value of the in-plane lattice parameter "$a$" (= |$\mathbf{a}$| = |$\mathbf{b}$|).

## RESULTS AND DISCUSSIONS

The calculated ground state in-plane lattice constants of unstrained 2D h-BN and h-AlN were respectively obtained as $a_0$(h-BN)= 2.488 Å and $a_0$(h-AlN) = 3.09 Å, which are in excellent agreement with the reported theoretical results of $a_0$(h-BN)= 2.488 Å in [13] and $a_0$(h-AlN)= 3.09 Å in [10]. As seen in Figure 1(a), 2D h-BN with its both valence-band maximum (VBM) and conduction band minimum (CBM) located at the K point of the Brillouin Zone (BZ), revealed a direct band gap of $E_g$(h-BN) = 4.606 eV, in agreement with previous calculations of 4.61 eV in [10] and 4.613 eV in [13]. Our estimated band gap energy of 2D h-BN is about 23% less than the experimental [6] direct band gap energy of 5.971 eV. This is due to the well known band gap (underestimation) problem within LDA. As seen in Figure 1(b), the band gap of 2D h-AlN is indirect with VBM located at K and CBM at Γ and has a value $E_g$(h-AlN) = 3.037 eV, which is in agreement with the reported [10] calculated value of 3.08 eV.

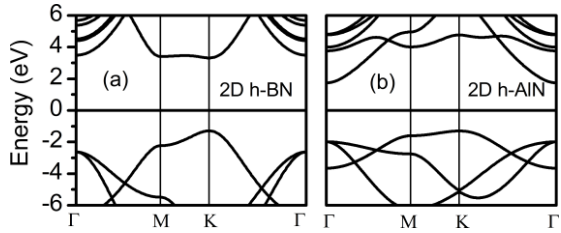

**Figure 1.** Energy bands of 2D h-BN (a) and h-AlN (b).

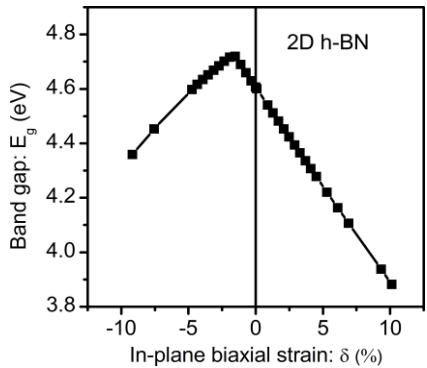

**Figure 2**. Variation of $E_g$ of 2D h-BN with in-plane homogeneous biaxial strain δ = $(a - a_0)/a_0$.

The nature of variations of our calculated values of $E_g$ with in-plane homogeneous biaxial strain δ = $(a - a_0)/a_0$ for 2D h-BN and h-AlN are respectively depicted in Figure 2 and Figure 3. For strain values in the range from -1.53% up to +10% the $E_g$(h-BN) remains direct; and for strains below -1.53% down to -10%, $E_g$(h-BN) remains indirect with VBM located at the K and CBM at the Γ. These results corroborate the reported results [13], and gives us confidence in our calculations, especially when applied to 2D h-AlN, for which we find a strong nonlinear variation of the band gap with strain as shown in Figure 3; the gap remains indirect within the considered strain range of ±10%, which is our new result.

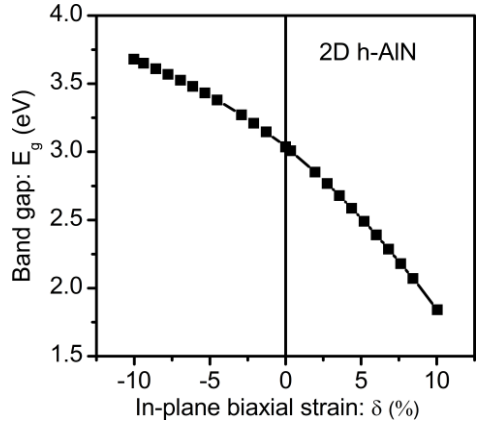

**Figure 3**. Variation of $E_g$ of 2D h-AlN with in-plane homogeneous biaxial strain δ = $(a - a_0)/a_0$.

## CONCLUSIONS

In this DFT based FP-(L)APW+lo study of the effects of biaxial strain on 2D h-BN and h-AlN, we found strain tunable energy band gaps of these nano-structures. While our result on h-BN corroborate the reported strain effects on h-BN, our result on strain tunable band gap of 2D-h-AlN is something new. These results may find applications in fabrication of NEMS based on h-BN and h-AlN.